\documentclass[doublespacing]{elsart}

\usepackage{natbib}

\usepackage{graphicx}

\usepackage{amsmath,amssymb}

\renewcommand{\cite}{\citep}

\begin{document}

\begin{frontmatter}
\title{Effect of density step on 
       stirring properties of a strain flow}
\author{M. Gonzalez\corauthref{cor1}}
\author{and}
\author{P. Parantho\"en}
\corauth[cor1]{Email address: Michel.Gonzalez@coria.fr}
\address{
CNRS, UMR 6614, Laboratoire de Thermodynamique, CORIA, \\
Site universitaire du Madrillet, 
76801 Saint-Etienne du Rouvray, France
}
\begin{abstract}
The influence of steep density gradient upon
stirring properties of a strain flow is addressed
by considering the problem in which an interface 
separating two regions with different constant densities
is stabilised within a stagnation-point flow.
The existence of an analytic solution for the 
two-dimensional incompressible flow field
allows the exact derivation of the velocity gradient
tensor and of parameters describing the local flow topology.
Stirring properties are affected not only through vorticity
production and jump of strain intensity at the interface,
but also through rotation of strain principal axes
resulting from anisotropy of pressure Hessian. 
The strain persistence parameter, which
measures the respective effects of strain and effective
rotation (vorticity plus rotation rate of strain basis),
reveals a complex structure.
In particular, 
for large values of the density ratio 
it indicates dominating effective rotation 
in a restricted area past the interface.
Information on flow structure
derived from
the Okubo-Weiss parameter, by contrast, is less detailed.
The influence of the density step on stirring properties
is assessed through the Lagrangian evolution of the gradient
of a passive scalar. Even for a moderate density ratio,
alignment of the scalar gradient  
and
growth rate of its norm
are deeply altered. Past the interface effective rotation
indeed drives the scalar gradient to align
with a direction
determined by the local strain persistence parameter, away
from the compressional strain direction. The jump of strain
intensity at the interface, however, opposes the
lessening
effect of the latter mechanism on the growth rate of the
scalar gradient
norm
and promotes the rise of the gradient.
\end{abstract}
\begin{keyword}
strain flow, density step, flow topology, stirring properties,
passive scalar gradient
\end{keyword}
\end{frontmatter}

\section{Introduction} \label{sec:introduction}

In stratified environmental 
and engineering flows or in reacting flows 
where
local
heat release 
occurs
velocity gradient properties are affected by
spatial variations of
density.
Density gradients may thus indirectly 
play on
stirring properties of flows through the local
features of the velocity gradient.
The 
influence of a density gradient upon velocity gradient properties
has been tackled in some studies.
In their classification of local flow topologies
\citet{Cal90} considered the case of compressible flows.
\citet{Hal98} have shown analytically 
how
in geostrophic turbulence
the topology of stirring 
undergoes the effect of
a stable stratification through the velocity
gradient.
The interaction between strain, vorticity and density
gradient in homogeneous 
sheared turbulence 
has been studied numerically by \citet{DN00}.
In combustion flows heat release may deeply influence mixing
through scalar dissipation \cite{Pal03,CS07}.
This finding thus suggests a possible effect of density variations 
on the velocity gradient/scalar gradient interaction.
In passing, 
it also questions the relevance to combustion
of `cold-flow analyses' of small-scale mixing.
The small-scale structure of the velocity field in nonpremixed
flames has been investigated in the numerical studies of
\citet{NE93} and \citet{Bal96,Bal98}.
These authors showed how strain and vorticity properties
-- in particular, vorticity alignments -- are changed by variable
density via dilatational and baroclinic effects. \citet{NE93}
also studied the local flow topology using
the classification of \citet{Cal90} for compressible flows.
\citet{Bal98} analysed the way in which buoyancy-influenced
vorticity alignment indirectly affects the dissipation rate
of fuel concentration. In addition, they examined how the properties 
of the pressure Hessian are changed by variable density.
This is a key point, for the pressure Hessian contributes to
the rotation of strain principal axes \cite{DT91,NP98,Lal99}
and may thus be indirectly involved in the small-scale mixing
process through alignment mechanisms.

The present study is specially focused on
the effect of local steep density gradients. 
The latter are produced, for instance,
in reacting flows with significant heat release. 
For the purpose of our analysis we
consider the flow field of a model problem taken from flame studies
in which
an interface stabilised in a pure straining velocity
field divides the flow in two regions with different densities.
The two dimensional inviscid problem in which each flow region is
assumed to be incompressible has an analytic solution
\cite{KM88}.
The main advantage of the approach lies in that
the
flow field is completely known and its 
properties in terms of strain, vorticity,
pressure Hessian {\ldots} can therefore be derived 
exactly.
In particular, parameters relevant to local flow topology
and stirring properties
such as the Okubo-Weiss parameter or the strain persistence 
parameter can be
expressed analytically. This makes it possible to
examine thoroughly the way in which they are influenced
by the density step.
Finally,
the strain persistence parameter is used to track the 
Lagrangian evolution of the gradient of a passive
scalar in terms of orientation and norm growth rate.
The effects of the density step upon the stirring
mechanism, then, are precisely revealed.

The problem under study and the relating flow field are 
described in Section \ref{sec:sec2}. The flow structure
derived from 
the velocity gradient tensor is studied
in Section \ref{sec:sec3}. 
In particular, the way in which the level of the density
step affects the local flow topology is closely analysed
through the behaviour of the strain persistence parameter.
In Section \ref{sec:sec4} 
we show how 
the local alignment 
and growth rate
of the gradient of a passive scalar and thus
stirring properties are influenced
by the density step. Conclusions are drawn in Section 
\ref{sec:conclusion}

\section{Flow conditions and dynamic field equations} \label{sec:sec2}

We consider the stationary problem analysed by \citet{KM88}
which consists of a plane premixed flame stabilised
in the stagnation-point flow
of a body
(Fig.~\ref{Fig:fig1}).
The 
flame thickness is assumed to be much smaller
than its standoff distance. At the flow scale, the flame front
may thus be regarded as
a discontinuity separating unburnt from lighter
burnt gases.
In addition, 
the flow is considered to be inviscid
and
the flowfield
in each region of constant density 
($ \rho_u $ and $ \rho_b $, respectively, in the unburnt
and burnt gases)
is described by the two-dimensional incompressible Euler equations.
Stability analysis has been made by \citet{KM90}
and \citet{JM93}. 
In the present study
stability is ensured according to the
criteria derived by \citet{KM90}
for the two-dimensional case, 
in particular, unity Lewis number is prescribed.

The equations are made dimensionless by using
the adiabatic flame speed 
and
the length scale characterising the standoff distance
as
velocity and length units;  
density, temperature and pressure
are normalised
by their respective values in the unburnt gas.
\citet{KM88}
derived
the velocity field from
Euler and continuity equations combined with Rankine-Hugoniot
relations for the jumps at the interface and boundary conditions 
for velocity 
upstream ($ x \rightarrow \infty $)
and
at the surface of the body ($ x = 0 $).
The flame is found to stabilise at distance $ d $ from
the body
which, for unity Lewis number and plane geometry, is given by:
\begin{equation}
d =
\frac{2 \gamma \arctan \left[{(\gamma - 1)}^{1/2} \right]}
     {\sigma_0 {(\gamma - 1)}^{1/2}}.
\label{d}
\end{equation}
In the case of non-unity Lewis number, $ d $ is given by a
transcendental equation which precludes a complete analytic study.
Parameters $ \gamma = \rho_u/\rho_b $ and $ \sigma_0 $ are, respectively,
the thermal expansion coefficient
(or density ratio)
and the intensity of the imposed  
strain.
Strain intensity is defined as
$ \sigma = {(\sigma_n^2 + \sigma_s^2)}^{1/2} $
where
$ \sigma_n = \partial u/\partial x - \partial v/\partial y $
and
$ \sigma_s = \partial v/\partial x + \partial u/\partial y $
are the normal and shear components of strain.
In the unburnt region $ \sigma = \sigma_0 $.
The coordinate system is defined in Fig.~\ref{Fig:fig1}
and $ u $ and $ v $ are the velocity components normal
and parallel to the interface, respectively.

The velocity and pressure fields are described by the following
equations \cite{KM88}.

For $ x > d $:
\begin{equation}
\rho = 1,
\label{rho1}
\end{equation}
\begin{equation}
u = -\frac{1}{2} \sigma_0 (x - a),
\label{u1}
\end{equation}
\begin{equation}
v=\frac{1}{2} \sigma_0 y,
\label{v1}
\end{equation}
\begin{equation}
p = p_a - \frac{1}{8} \sigma_0^2 \left[{(x -a)}^2 + y^2\right],
\label{p1}
\end{equation}

where
$ p_a $ 
is
the pressure at the virtual stagnation point, $ (x,y) = (a,0) $,
and
constant $ a $ is given by:
\begin{equation}
a =
\left[
1 - \frac{{(\gamma - 1)}^{1/2}}
         {\gamma \arctan \left[{(\gamma - 1)}^{1/2} \right]}
\right] d.
\label{a}
\end{equation}

For $ x < d $:
\begin{equation}
\rho = \frac{1}{\gamma},
\label{rho2}
\end{equation}
\begin{equation}
u = -\frac{1}{2} \sigma_0 \delta \sin (k x),
\label{u2}
\end{equation}
\begin{equation}
v = \frac{1}{2} \sigma_0 k \delta y \cos (k x),
\label{v2}
\end{equation}
\begin{equation}
p = p_s - \frac{1}{8} \sigma_0^2
\left[
\frac{\delta^2}{\gamma} \sin^2 (k x)
+
y^2
\right],
\label{p2}
\end{equation}

with
pressure at the actual stagnation point, $ p_s $, given by:
\begin{equation}
p_s =
p_a - \frac{1}{8} (\gamma - 1) {(d - a)}^2 \sigma_0^2,
\label{ps}
\end{equation}

and constants $ \delta $ and $ k $ by:
\begin{equation}
\delta =
\gamma {\left(
\frac{\gamma}{\gamma - 1}
        \right)}^{1/2}
(d - a),
\label{delta}
\end{equation}

and
\begin{equation}
k =
\frac{{(\gamma - 1)}^{1/2}}
     {\gamma (d - a)}.
\label{k}
\end{equation}

The dynamic field is completely defined by
(\ref{d}) to (\ref{k})
and parameters $ \gamma $ and $ \sigma_0 $.
Note that from (\ref{d}) and (\ref{a})
$ d - a $ only depends on the imposed strain as
$ d - a = 2/\sigma_0 $.
In the following we use the above model for 
deriving
the change in flow structure 
brought about by the density step.
We are more specifically interested in 
stirring properties resulting from the flow
topology
determined by
the
velocity gradient tensor.

\section{Influence of the density step on flow structure} \label{sec:sec3}

The velocity gradient tensor can be 
derived analytically from 
(\ref{d}) -- (\ref{k}).
In particular, calculation of its symmetric
and antisymmetric parts, namely strain and rotation
is straightforward.

The properties of the prescribed strain flow 
($ x > d $)
are obviously those
of a pure straining motion with $ \sigma_n = -\sigma_0 $
and $ \sigma_s = 0 $;
the Okubo-Weiss parameter,
$ Q = \sigma^2 - \omega^2 $, where 
$ \omega = \partial v/\partial x - \partial u/\partial y $ is
the vorticity,
amounts to $ Q = \sigma_0^2 $.
The pressure Hessian, $ \boldsymbol{H} $, is isotropic; 
from (\ref{p1}):
\begin{equation}
H_{ij} = \frac{1}{\rho}\frac{\partial^2 p}{\partial x_i \partial x_j}
       =-\frac{1}{4} \sigma_0^2 \delta_{ij}.
\label{Hij1}
\end{equation}
The strain persistence parameter, $ r $, is defined as
\cite{Lal99}:
\begin{equation}
r
=
\frac{\omega + \Omega}{\sigma},
\label{r}
\end{equation}
where
$ \Omega $ is the rotation rate of strain
principal axes \cite{Lal99}:
\begin{equation}
\Omega
=
\frac{1}{\sigma^2}
\left[
\sigma_s \left(\frac{d \sigma_n}{d t}\right)
-
\sigma_n \left(\frac{d \sigma_s}{d t}\right)
\right].
\label{Omega}
\end{equation}
Because it includes 
the acceleration gradient
tensor
via
the strain derivatives, 
the strain persistence parameter is more
general than the Okubo-Weiss parameter.
In particular,
accounting for 
the Lagrangian variation of strain allows 
a better 
estimate of stirring properties \cite{HK98,Lal99}.
Dominating strain corresponds to $ r^2 < 1 $, while 
effective rotation prevails if $ r^2 > 1 $   
and balance occurs for $ r^2 = 1 $.
Analysis of stirring through the behaviour of a passive scalar gradient
in terms of orientation and norm based on the latter different
regimes
has been
performed
by \citet{Lal99}.
They 
assumed
$ r $  
varying ``slowly'' 
along the Lagrangian trajectories, that is, on a time scale
much larger than the gradient response time scale.
The effect of fluctuations of $ r $
faster than the response of the scalar gradient
has been studied by \citet{Gal05,Gal08}.
In  
an
inviscid incompressible flow
equations for $ \sigma_n $ and $ \sigma_s $
only
include the pressure Hessian components
\cite{Lal99}:
\begin{equation}
\frac{d \sigma_n}{d t}
=
\frac{1}{\rho} \frac{\partial^2 p}{\partial y^2}
-
\frac{1}{\rho} \frac{\partial^2 p}{\partial x^2},
\label{eqsigman}
\end{equation}
\begin{equation}
\frac{d \sigma_s}{d t}
=
- \frac{2}{\rho} \frac{\partial^2 p}{\partial x \partial y},
\label{eqsigmas}
\end{equation}
and reveal that 
the rotation rate of strain axes, $ \Omega $, results
from the anisotropic part of $ \boldsymbol{H} $.
Equations (\ref{Hij1}) and (\ref{Omega})    
to (\ref{eqsigmas}) thus 
show that in the imposed straining flow $ \Omega = 0 $
and, as a consequence, $ r = 0 $ which indicates perfectly
persistent strain. 
The Okubo-Weiss and strain persistence parameters
obviously agree in this simple case. 
However,  depending on the density ratio, they may differ
from one another in their diagnoses in the flow region past
the interface as shown
in the following.

Beyond the interface ($ x < d $)
the velocity gradient has a rotational part
due to vorticity production at the 
discontinuity.
Strain and vorticity
are derived from
(\ref{u2}), (\ref{v2}),
(\ref{delta}) and (\ref{k}):
\begin{equation}
\sigma_n
=
-\sigma_0 \gamma^{1/2} \cos (k x),
\label{sigman2}
\end{equation}
\begin{equation}
\sigma_s
=
- \frac{1}{2} \sigma_0 \gamma^{1/2} k y \sin (k x),
\label{sigmas2}
\end{equation}
\begin{equation}
\omega
=
- \frac{1}{2} \sigma_0 \gamma^{1/2} k y \sin (k x).
\label{omega2}
\end{equation}
At the interface, the normal component of strain is conserved,
$ \sigma_n(d^-) = \sigma_n(d^+) = -\sigma_0 $, while the
shear component undergoes a step from $ \sigma_s(d^+) = 0 $ to
$ \sigma_s(d^-) = -\sigma_0 {(\gamma - 1)}^{1/2}
  \arctan\left[{(\gamma - 1)}^{1/2}\right] y/2d $ resulting in a step in
strain intensity expressed by:
\begin{equation}
\sigma(d^-)
=
\sigma(d^+){\left[
1 + (\gamma - 1) \arctan^2\left[{(\gamma - 1)}^{1/2}\right]
\frac{y^2}{4 d^2} \right]}^{1/2}
\label{sigmastep}
\end{equation}
The Okubo-Weiss parameter depends on both imposed strain
and density ratio and varies with
$ x $
as:
\[
Q
=
\sigma_0^2 \gamma \cos^2 (k x).
\]
Because from (\ref{d}) and (\ref{k}) 
$ k x \leq k d = \arctan \left[{(\gamma -1)}^{1/2} \right] < \pi/2 $,
parameter
$ Q $ never takes a zero value.
It
grows with increasing density ratio and is
strictly positive thus indicating
prevailing
strain.

The pressure Hessian derived from (\ref{rho2}) and (\ref{p2})
(using also (\ref{delta}) and (\ref{k})) is
anisotropic: 
\begin{equation}
\frac{1}{\rho} \frac{\partial^2 p}{\partial x^2}
=
- \frac{1}{4} \sigma_0^2 \gamma \cos (2 k x),
\label{H112}
\end{equation}
\begin{equation}
\frac{1}{\rho} \frac{\partial^2 p}{\partial y^2}
=
- \frac{1}{4} \sigma_0^2 \gamma, 
\label{H222}
\end{equation}
\begin{equation}
\frac{1}{\rho} \frac{\partial^2 p}{\partial x \partial y}
=
\frac{1}{\rho} \frac{\partial^2 p}{\partial y \partial x}
= 0.
\label{H122}
\end{equation}
From (\ref{k}) and (\ref{H112}) it is clear that 
anisotropy is caused by the density step in $ x $ direction
(which was foreseeable)
and
that isotropy of the pressure
Hessian is retrieved for $ \gamma = 1 $.
The level of anisotropy of $ \boldsymbol{H} $ is given by
the anisotropy tensor,
$ \boldsymbol{A} $,
defined by:
\begin{equation}
A_{ij}
=
\frac{2 H_{ij} - H_{\alpha \alpha} \delta_{ij}}
     {H_{\alpha \alpha}}.
\label{Aij}
\end{equation}
From (\ref{H112}) -- (\ref{Aij}):
\[
A_{11}
=
- A_{22}
=
- \tan^2 (k x), 
\]
and
\[
A_{12}
=
A_{21}
= 0.
\]
The norm of tensor $ \boldsymbol{A} $,
$ | \boldsymbol{A} | = \sqrt 2 \tan^2 (k x) $,
is plotted in Fig.~\ref{Fig:fig2}
for different values of $ \gamma $.
As $ \gamma $ is increased,
the level of anisotropy 
grows and
the largest anisotropy is restricted to a region
closer to the interface.
Defining $ x^{\star}_{1/2} $ as the location in $ x/d $
where $ | \boldsymbol{A} | $ reaches half its maximum
value,
it is straightforward to show that
\[
x^{\star}_{1/2}
=
\frac{\arctan \left[{\left(\frac{\gamma-1}{2}\right)}^{1/2} \right]}
     {\arctan \left[{(\gamma - 1)}^{1/2} \right]},
\]
and, then, $ x^{\star}_{1/2} $
tends to unity with increasing $ \gamma $.
Note that plotting results (possibly normalised by the imposed strain,
$ \sigma_0 $, when needed) in function of $ x/d $, $ y/d $ or $ \sigma_0 t $
(for Lagrangian tracking) 
allows to get rid of explicit dependence on $ \sigma_0 $ and 
to
restrict 
the analysis
to the effects of 
density ratio.

The anisotropy of the pressure Hessian results in rotation
of the strain principal axes. Using (\ref{Omega}) 
with 
(\ref{eqsigman}) -- (\ref{sigmas2}) and the components of the
pressure Hessian given by 
(\ref{H112}) -- (\ref{H122}):
\begin{equation}
\Omega
=
\frac{\sigma_0 \gamma^{1/2} k y \sin (kx) \tan^2 (kx)}
     {4 + k^2 y^2 \tan^2 (kx)}.
\label{Omega2}
\end{equation}
It is worth noticing that 
from their respective signs 
the rotation rate of strain basis,
$ \Omega $ and vorticity, $ \omega $ (given by (\ref{omega2})),
oppose each other.

The strain persistence parameter, then, is derived from
(\ref{r}) together with (\ref{sigman2}) -- (\ref{omega2})
and (\ref{Omega2}):
\[
r
=
-
\frac{[4 + (k^2 y^2 - 2) \tan^2 (kx)] k y \tan (kx)}
     {{[4  + k^2 y^2 \tan^2 (kx)]}^{3/2}}.
\]
Function $ r $ depends on density ratio through $ k $
given by (\ref{k}).
It is antisymmetric in $ y $ and tends to -1 from above as
$ y $ tends toward $ + \infty $ and to 1 from below as $ y $
tends toward $ - \infty $.
A closer analytic examination
reveals a rather rich behaviour
that depends on the value of the density ratio.
For $ \gamma \leq 3 $, $ r $ is a monotonous decreasing function
as shown in Fig.~\ref{Fig:fig3}.
For $ 3 < \gamma \leq 9 $, extrema appear 
near the interface, in a region 
defined by
$ (x/d)_1 < x/d \leq 1 $ (Fig.~\ref{Fig:fig4}). 
Finally,
if $ \gamma > 9 $, $ r $ has a three-layered structure in 
$ x $; $ x/d \leq (x/d)_1 $: $ r $ is monotonous and decreasing;
$ (x/d)_1 < x/d \leq (x/d)_2 $: $ r $ has extrema within 
the bounds -1 and 1; $ (x/d)_2 < x/d $: the extrema excede
-1 and 1. This behaviour is displayed in Fig.~\ref{Fig:fig5}. 
The threshold values $ (x/d)_1 $ and $ (x/d)_2 $ are given by:
\[
{\left(     
      \frac{x}{d}
\right)}_1
=
\frac{\arctan \sqrt 2}
     {\arctan \left[{(\gamma - 1)}^{1/2} \right]},
\]
and
\[
{\left(     
      \frac{x}{d}
\right)}_2
=
\frac{\arctan(2 \sqrt 2)}
     {\arctan \left[{(\gamma - 1)}^{1/2} \right]},
\]
and,
as $ \gamma $ is increased,
tend to constant values,
namely
$ (2 \arctan \sqrt 2) / \pi \simeq 0.608 $
and
$ [2 \arctan (2 \sqrt 2)] / \pi \simeq 0.784 $.
Figure~\ref{Fig:fig6} shows the two-dimensional field of 
$ r^2 $ 
derived for a large value of the density ratio,
$ \gamma = 10 $.
Beyond the interface strain significantly dominates
only
in the vicinity of the plane of symmetry and of the slip
boundary located at $ x/d =0 $.

The evolution of $ r $ at large density ratio is explained
by the rotation of strain principal axes. The vorticity-to-strain
ratio defined as $ R = \omega/\sigma $ discarding the 
rotation rate of strain basis, $ \Omega $, has a simpler
behaviour.
From (\ref{sigman2}) -- (\ref{omega2}):
\[
R
=
- \frac{k y \tan(kx)}
       {{[4 + k^2 y^2 \tan^2(kx)]}^{1/2}},
\]
which shows that  $ R $ is monotonous
and bounded by -1 and 1
no matter the value of the density ratio. This is in no way
surprising, for the criterion based on $ R $  
only includes vorticity and strain and therefore
amounts
to the Okubo-Weiss criterion.
Figures~\ref{Fig:fig7} and ~\ref{Fig:fig8} display
profiles of $ R $ for $ \gamma =2 $ and $ \gamma = 10 $,
respectively.
Better still,
from (\ref{r}) and the fact that from
(\ref{omega2}) and ({\ref{Omega2}) 
$ \omega $ and $ \Omega $ have opposed signs,
$ r $ starts developing a positive 
(resp. negative)
extremum for $ y > 0 $
(resp. $ y < 0 $),
near the origin, provided that 
$ | \Omega/\omega | > 1 $.
From the exact expression for $ \Omega/\omega $ derived
using (\ref{omega2}) and (\ref{Omega2}), it can be checked
that $ | \Omega/\omega | > 1 $ for $ \gamma > 3 $.
The region where $ | \Omega/\omega | > 1 $ is found to
coincide with the one where $ r $ displays extrema.
It is also found that in the present case $ | r | > 1 $
for $ | \Omega/\omega | > 4 $.

Since
in the problem under study the solution given by
(\ref{u1}) -- (\ref{a}) and (\ref{u2}) -- (\ref{k})
depends on Lewis number through distance $ d $
\cite{KM88}, the above threshold values for $ \gamma $
and $ x/d $ 
derived in the case of unity Lewis number are not universal.
Nevertheless,
the analysis leads to the conclusion that
past the density interface
parameter $ r $ indicates
dominating strain
except
when the density ratio is large enough;
in this case, effective rotation prevails 
near the plane of symmetry, in a region close
to the interface.

\section{Consequences for alignment and stirring properties}\label{sec:sec4}

Stirring and mixing
properties of a flow field can be analysed through the
evolution of the gradient of an advected passive scalar. In this
view 
alignment of the scalar gradient with respect to the local strain
principal axes 
is a key mechanism
\cite{P94,Val01,Bal03}.
The increase of the gradient norm brought about by
alignment with the local compressional direction
causes
enhancement 
of mixing through acceleration of molecular diffusion.

\citet{Lal99} have shown that in two-dimensional flows 
the opposed effects of strain and effective rotation
result in the existence of directions that are mostly
different from those of strain principal axes.
Their analysis is based on the equation  
in the strain basis
for the orientation 
of the gradient of a non-diffusive passive scalar:
\begin{equation}
\frac{d \zeta}{d \tau}
=
r - \cos \zeta.
\label{zeta}
\end{equation}
In the fixed frame of reference the scalar gradient is defined
by vector
$ \boldsymbol{G} = |\boldsymbol{G}| (\cos \theta, \sin \theta) $
with
$ \zeta = 2 (\theta + \Phi) $  
and $ \Phi $ gives the orientation of the strain principal
axes through $ \tan (2 \Phi) = \sigma_n /\sigma_s $.
Note that the rotation rate, $ \Omega $, of the strain principal axes 
defined by (\ref{Omega}) coincides with $ 2 d \Phi/d t $.
Time $ \tau $ is a strain normalised time:
\[
\tau
=
\int_0^t \sigma(t') dt',
\]
where $ t $ stands for the Lagrangian time
and $ \sigma $ for strain intensity.
Assuming slow variations of $ r $ along Lagrangian trajectories,
\citet{Lal99} showed how the local flow topology determines
the orientation of the scalar gradient. If strain prevails
over effective rotation ($ r^2 < 1 $), the orientation equation
(\ref{zeta}) has a stable fixed point,
\begin{equation}
\zeta_{eq}(r)
=
- \arccos r,
\label{zetaeq}
\end{equation}
corresponding to an equilibrium orientation.
It is only in the special case $ r = 0 $ 
-- i.e. in the pure hyperbolic regime --
that the equilibrium
orientation, $ \zeta_{eq} $, coincides with the local
compressional direction,
$ \zeta_c = - \pi/2 $,
which corresponds to
$ \theta = - \Phi -\pi/4 $ in the fixed
frame of reference.
When strain and effective rotation balance each other ($ r^2 =1 $),
the equilibrium orientation is a bisector of strain
principal axes.
If
effective rotation dominates ($ r^2 > 1 $), the scalar has no equilibrium
orientation, 
but a most probable one 
coinciding with a bisector of the strain basis.
From the numerical simulations of \citet{Lal99} in two-dimensional
turbulence it appears that 
in strain-dominated regions
the scalar gradient 
statistically aligns better with the local equilibrium orientation
than with the compressional direction.
In fact,
\citet{Gal05,Gal08} have shown that the scalar 
gradient aligns with the equilibrium direction rather
than with the compressional one provided that its response
time scale to $ r $ fluctuations is short enough compared
to the Lagrangian time scale of $ r $.

In the above approach the equation for the norm of 
the scalar gradient is \cite[]{Lal99}:
\[
\frac{2}{|\boldsymbol{G}|}
\frac{d |\boldsymbol{G}|}{d t}
=
- \sigma \sin \zeta,
\]
and obviously defines the local growth rate of the gradient norm as
$ - \sigma \sin \zeta $.
Clearly, alignment with the compressional direction,
$ \zeta = \zeta_c = - \pi/2 $, corresponds to the maximum 
growth rate.
The growth rate in the unaltered region, $ x > d $, obviously
amounts to $ \sigma_0 $.

The local stirring properties of the flow under study
can be derived from the field of the strain persistence
parameter, $ r $. In particular, the local equilibrium
orientation is known through (\ref{zetaeq}).
Figure~\ref{Fig:fig9} displays the field of variable
$ \Delta \zeta =
|2 [\zeta_{eq}(r) + \pi/2]/\pi| $
in the case of density ratio $ \gamma = 6 $.
For the latter value of $ \gamma $ the whole flow is
dominated by strain and $ \zeta_{eq} $ is defined everywhere.
Variable $ \Delta \zeta $
ranges from 0 to 1 and
gives the normalised difference between the local equilibrium
and
compressional directions. 
It takes the maximum value, $ \Delta \zeta = 1 $,
for $ \zeta_{eq} $ corresponding to a bisector 
of the strain principal axes, namely $ \zeta_{eq} = 0 $
or $ \zeta_{eq} = -\pi $ which coincide, respectively, with
directions  
$ - \Phi $ and $ - \Phi -\pi/2 $ in the fixed frame
of reference.   
From Fig.~\ref{Fig:fig9} it is clear that
past the interface
$ \zeta_{eq} $
significantly departs from $ \zeta_c $ over most of the flow
field except 
near the plane of symmetry and the slip boundary
where strain significantly prevails over effective rotation.
For $ x/d > 1 $ pure strain 
makes the equilibrium orientation coincide
with the compressional direction.
According to the analysis of function $ r $ we made in
Section \ref{sec:sec3}, the difference $ \Delta \zeta $ obviously tends to
its maximum value 
with increasing $ y/d $.

The limitation of
stirring
properties 
which would result from
alignment with the local equilibrium 
orientation
instead of the compressional direction
is
given by the theoretical 
growth
rate,
normalised by local strain,
$ \eta_{eq}/\sigma = -\sin[\zeta_{eq}(r)]  
= \eta_{eq}^{\star} $ 
(Fig.~\ref{Fig:fig10}).
As expected, $ \eta_{eq}^{\star} $ is close to its maximum value near the 
plane of symmetry and the slip boundary
where the equilibrium and compressional directions 
almost coincide.
However, $ \eta_{eq}^{\star} $ displays lower values over the rest of 
the flow field and even tends to 0 for large values
of $ y/d $.
This trend results from 
$ \zeta_{eq} $
drawing near
values 0 or $ -\pi $ 
as $ r $ tends toward 1 or -1,
respectively.
But the step in strain intensity
given by (\ref{sigmastep}) 
overcomes the effect of possible misalignment of
the scalar gradient with respect to the compressional
direction. Figure ~\ref{Fig:fig11} shows the field of 
theoretical growth rate 
normalised by the imposed strain,
$ \eta_{eq}/\sigma_0
= \sigma \sin[\zeta_{eq}(r)]/\sigma_0 $,
and reveals the gain in
growth rate at the interface caused by the step in strain
intensity.

The response of a passive
scalar gradient to the variations of strain persistence along
Lagrangian trajectories 
brings further information.
Equation (\ref{zeta}) 
has been solved starting from initial
conditions at the interface ($ x/d = 1 $) for different values 
of $ y/d $, namely $ y(0)/d = $ 1, 2, 5 and 10
and density ratio $ \gamma = 6 $.
Figure \ref{Fig:fig12}
shows the corresponding trajectories (cut at $ y/d = 100 $).
Initial conditions for velocity are given by
(\ref{u2}) and (\ref{v2})
with $ x = d $.
Initially, the value of strain intensity 
is the imposed strain, $ \sigma_0 $,
strain persistence, $ r $, is zero and scalar gradient orientation
is $ \zeta = \zeta_c = -\pi/2 $.

Figure \ref{Fig:fig13} displays the trend of strain persistence
toward its zero, pure hyperbolic value after the sharp 
change
at the interface.
The larger $ y(0)/d $,
the deeper the variation caused by the interface and
the slower the evolution of $ r $;
at $ \sigma_0 t = 100 $ $ |r| $ is close to
0.1 if $ y(0)/d = 1 $, but is around 0.7
in the case $ y(0)/d = 10 $.
As expected, the local equilibrium orientation displays
a similar behaviour (Fig. \ref{Fig:fig14}):
for $ y(0)/d = 1 $
$ \zeta_{eq} $ almost retrieves its initial value, 
$ \zeta_c $, at
$ \sigma_0 t = 100 $; for $ y(0)/d = 10 $ 
$ \zeta_c - \zeta_{eq} $ is still greater than
$ 0.2 \pi $.
Now, the local, instantaneous orientation of the
scalar gradient coincides with $ \zeta_{eq} $ only
if its response time scale is short enough as     
compared to the time scale of 
$ r $ variations \cite{Gal05,Gal08}.
Figure \ref{Fig:fig15} shows 
that 
after $ \sigma_0 t = 2 $ this condition is realized
and $ \zeta $ varies as $ \zeta_{eq} $
no matter the value of $ y(0)/d $.
The actual growth rate of the scalar gradient norm thus
coincides with
the theoretical growth rate (Fig. \ref{Fig:fig11})
over most of the field past the interface and, as
shown in Fig. \ref{Fig:fig16}, 
reveals a significant change of stirring properties
with increasing $ y(0)/d $.
For long times, $ \zeta $ tends to $ \zeta_{eq} = -\arccos r $,
with $ r $ tending to 0; in addition, 
from (\ref{sigman2}) and (\ref{sigmas2}) strain
intensity, $ \sigma $, tends to $ \gamma^{1/2} \sigma_0 $ with $ x $
tending to 0.   
This makes the actual normalised growth rate,
$ \eta/\sigma_0 = - \sigma \sin \zeta/\sigma_0 $,
tends to $ \gamma^{1/2} $.
At the interface, $ \zeta $ is close to $ \zeta_c $ and
the jump of $ \eta/\sigma_0 $ coincides with the jump
of strain intensity given by (\ref{sigmastep}) and is of the
order of $ {(\gamma - 1)}^{1/2} \arctan\left[{(\gamma - 1)}^{1/2}\right] y/2d $.

\section{Conclusion} \label{sec:conclusion}

The flow field resulting from stabilisation of an infinitely
thin flame within a stagnation-point flow has been used 
for studying analytically 
the effect of a density step upon the structure and
stirring
properties of a straining flow.
The analysis rests on
the velocity gradient tensor and subsequent parameters indicating
the local flow
topology.

In addition to vorticity production and rise of strain
intensity, the density step causes rotation of strain
principal axes through anisotropy of the pressure Hessian.
Strain basis rotation and vorticity 
-- the sum of which corresponds to effective rotation --
do not combine, but oppose each other.
The Okubo-Weiss parameter, which is only based on strain
and vorticity,
is sensitive to the density step
and
indicates prevailing strain whatever the value of the
density ratio.
Because it accounts for the acceleration gradient through
the rotation rate of the strain basis, the strain persistence
parameter betrays a more subtle behaviour.
More specifically, 
it reveals the existence
of a restricted area past the interface where 
strain is reduced relatively to effective rotation
as
density ratio
is increased.
For large values of the density ratio, 
strain basis rotation overcomes vorticity and
effective rotation
prevails over strain in this region of the flow.

Moreover, alignment properties of the gradient of a passive scalar
and relating stirring properties of the flow field
can be derived from the strain persistence parameter.
For a moderate value of the density ratio, 
the difference between
the equilibrium orientation of the scalar gradient
resulting from the opposed effects of strain and effective 
rotation and the compressional
direction is significant over most of the flow field beyond the interface.
This misalignment with respect to compressional direction
tends to lessen the growth rate of the gradient norm, but
is overcome by the rise of strain intensity.
Lagrangian analysis 
shows that past the interface the local, instantaneous orientation
of the gradient of a passive scalar rapidly coincides
with the equilibrium orientation 
and confirms
the change in gradient growth rate (hence in stirring properties)
brought
about by the density step.

In this two-dimensional inviscid flow 
rotation of the strain
principal axes
is only caused
by anisotropy of the pressure Hessian,
while 
in the three-dimensional case 
it 
also results from local vorticity.
It would therefore be worth defining a three-dimensional
flow in which the question could be re-examined.
The strain/vorticity interaction also makes the three-dimensional
case more complex.
Another extension of the work should consist in relaxing
the assumption of an infinitely thin density interface
by defining a finite length scale of the density gradient.
The analytic approach would certainly be harder, but numerical
computation could be used for this purpose.

%


\newpage

\centerline{FIGURE CAPTION}

\bigskip

\noindent
FIGURE 1.
Stagnation-point flow 
altered
by the density step.
Streamlines are plotted for
density ratio
$ \gamma = 6 $.

\bigskip

\noindent
FIGURE 2.
Effect of density ratio on 
the norm of the anisotropy tensor
of pressure Hessian.

\bigskip

\noindent
FIGURE 3.
Profiles of strain persistence parameter for density ratio
$ \gamma = 2 $. 
Solid: $ x/d = 0.25 $; dashed: $ x/d = 0.50 $;
dashdot: $ x/d = 0.75 $; dotted: $ x/d = 1.0 $.

\bigskip

\noindent
FIGURE 4.
Profiles of strain persistence parameter for density ratio
$ \gamma = 6 $. 
Solid: $ x/d = 0.25 $; dashed: $ x/d = 0.50 $;
dashdot: $ x/d = 0.75 $; dotted: $ x/d = 0.90 $;
longdash: $ x/d = 1.0 $.

\bigskip

\noindent
FIGURE 5.
Profiles of strain persistence parameter for density ratio
$ \gamma = 10 $. 
Solid: $ x/d = 0.25 $; dashed: $ x/d = 0.50 $;
dashdot: $ x/d = 0.75 $; dotted: $ x/d = 0.90 $;
longdash: $ x/d = 1.0 $.

\bigskip

\noindent
FIGURE 6.
Field of strain persistence parameter 
-- squared --
for density ratio
$ \gamma = 10 $.

\bigskip

\noindent
FIGURE 7.
Profiles of vorticity-to-strain ratio for density ratio
$ \gamma = 2 $. 
Solid: $ x/d = 0.25 $; dashed: $ x/d = 0.50 $;
dashdot: $ x/d = 0.75 $; dotted: $ x/d = 1.0 $.

\bigskip

\noindent
FIGURE 8.
Profiles of vorticity-to-strain ratio for density ratio
$ \gamma = 10 $. 
Solid: $ x/d = 0.25 $; dashed: $ x/d = 0.50 $;
dashdot: $ x/d = 0.75 $; dotted: $ x/d = 1.0 $.

\bigskip

\noindent
FIGURE 9.
Normalised difference between the local compressional
direction and the equilibrium orientation of the gradient
of a passive scalar for density ratio $ \gamma = 6 $.

\bigskip

\noindent
FIGURE 10.
Theoretical growth rate of the norm of a passive scalar
gradient 
normalised by local strain;
density ratio $ \gamma = 6 $.

\bigskip

\noindent
FIGURE 11.
Theoretical growth rate of the norm of a passive scalar
gradient 
normalised by the imposed strain;
density ratio $ \gamma = 6 $.

\bigskip

\noindent
FIGURE 12.
Trajectories starting from $ x/d = 1 $
and $ y/d = $ 1, 2, 5  and 10; density ratio $ \gamma = 6 $.

\noindent
FIGURE 13.
Evolution of strain persistence parameter
along trajectories starting from $ x/d =1 $
and $ y/d = $ 1, 2, 5  and 10; density ratio $ \gamma = 6 $.

\noindent
FIGURE 14.
Evolution of equilibrium orientation
along trajectories starting from $ x/d =1 $
and $ y/d = $ 1, 2, 5  and 10; density ratio $ \gamma = 6 $.

\noindent
FIGURE 15.
Departure of scalar gradient orientation to
equilibrium orientation
along trajectories starting from $ x/d =1 $
and $ y/d = $ 1, 2, 5  and 10; density ratio $ \gamma = 6 $.

\noindent
FIGURE 16.
Evolution of the actual growth rate of scalar gradient norm
normalised by the imposed strain
along trajectories starting from $ x/d =1 $
and $ y/d = $ 1, 2, 5  and 10; density ratio $ \gamma = 6 $.

\newpage

\begin{figure}[htpb]
\begin{center}
\scalebox{0.5}{\includegraphics[angle=-90]{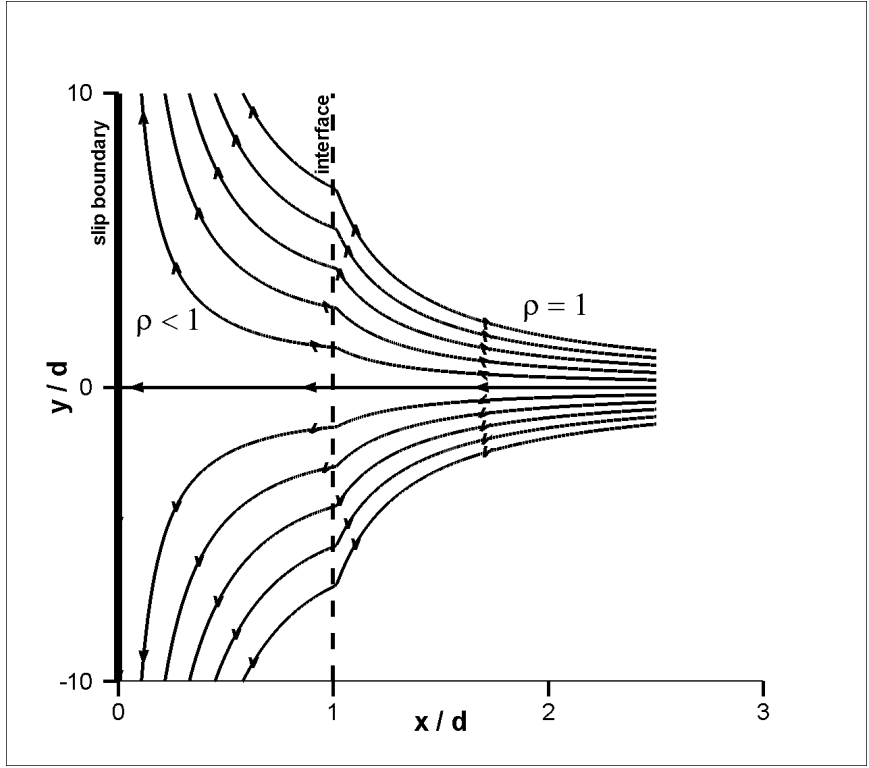}}
\caption{Stagnation-point flow altered by the density
         step. Streamlines are plotted for density
         ratio $ \gamma = 6 $.}
\label{Fig:fig1}
\end{center}
\end{figure}

\newpage

\begin{figure}[htpb]
\begin{center}
\scalebox{0.5}{\includegraphics[angle=-90]{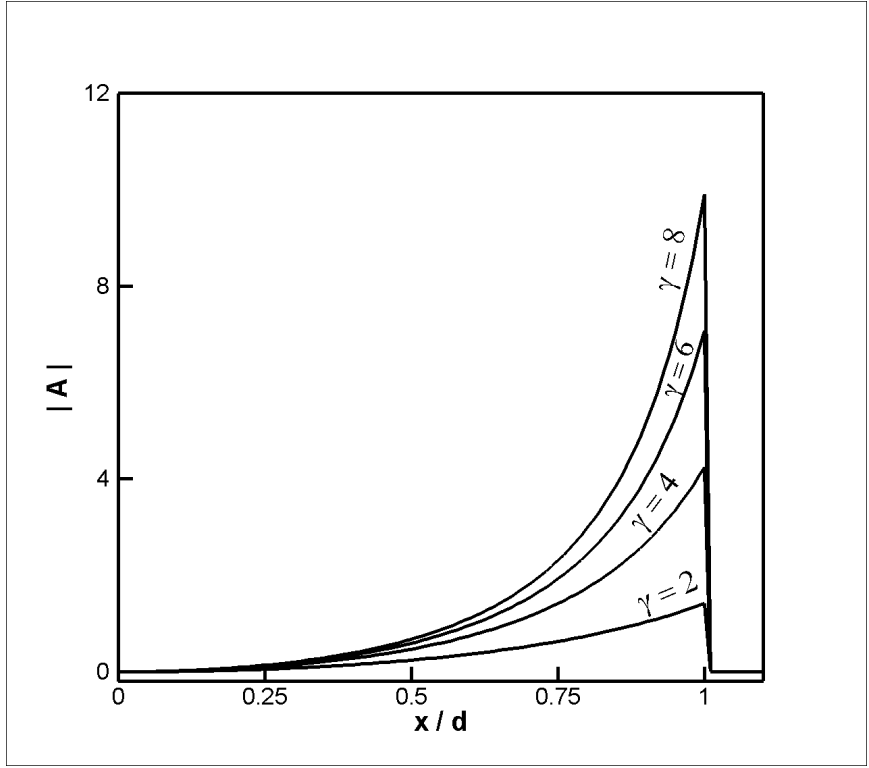}}
\caption{Effect of density ratio on the norm of 
         the anisotropy tensor of pressure Hessian.}
\label{Fig:fig2}
\end{center}
\end{figure}

\newpage

\begin{figure}[htpb]
\begin{center}
\scalebox{0.5}{\includegraphics[angle=-90]{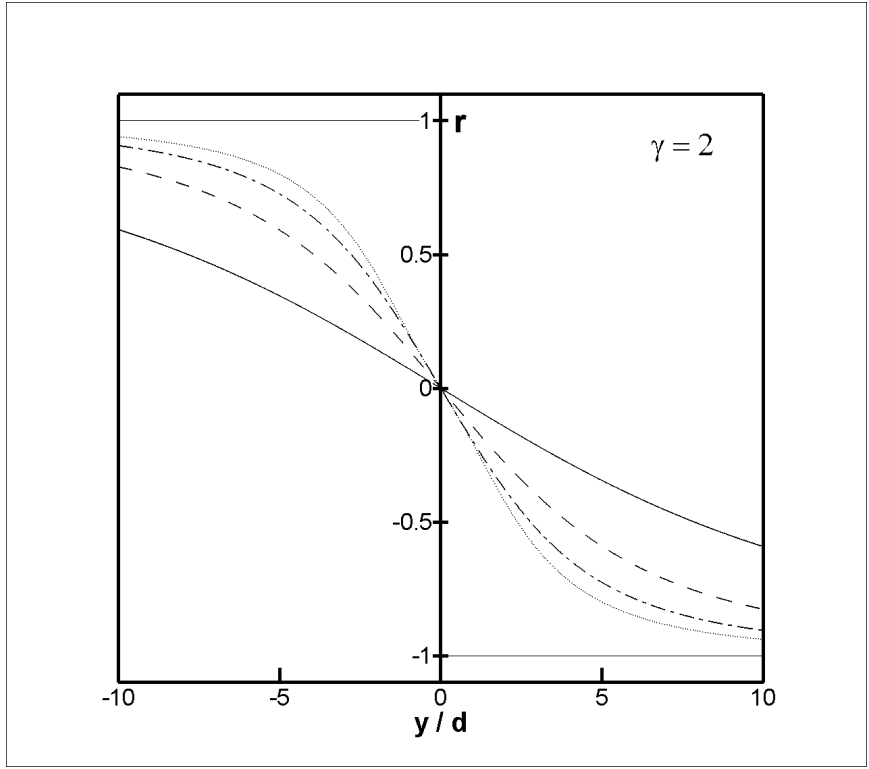}}
\caption{Profiles of strain persistence parameter for density
         ratio $ \gamma = 2 $.
         Solid: $ x/d = 0.25 $;
         dashed: $ x/d = 0.50 $;
         dashdot: $ x/d = 0.75 $;
         dotted: $ x/d = 1.0 $.}
\label{Fig:fig3}
\end{center}
\end{figure}

\newpage

\begin{figure}[htpb]
\begin{center}
\scalebox{0.5}{\includegraphics[angle=-90]{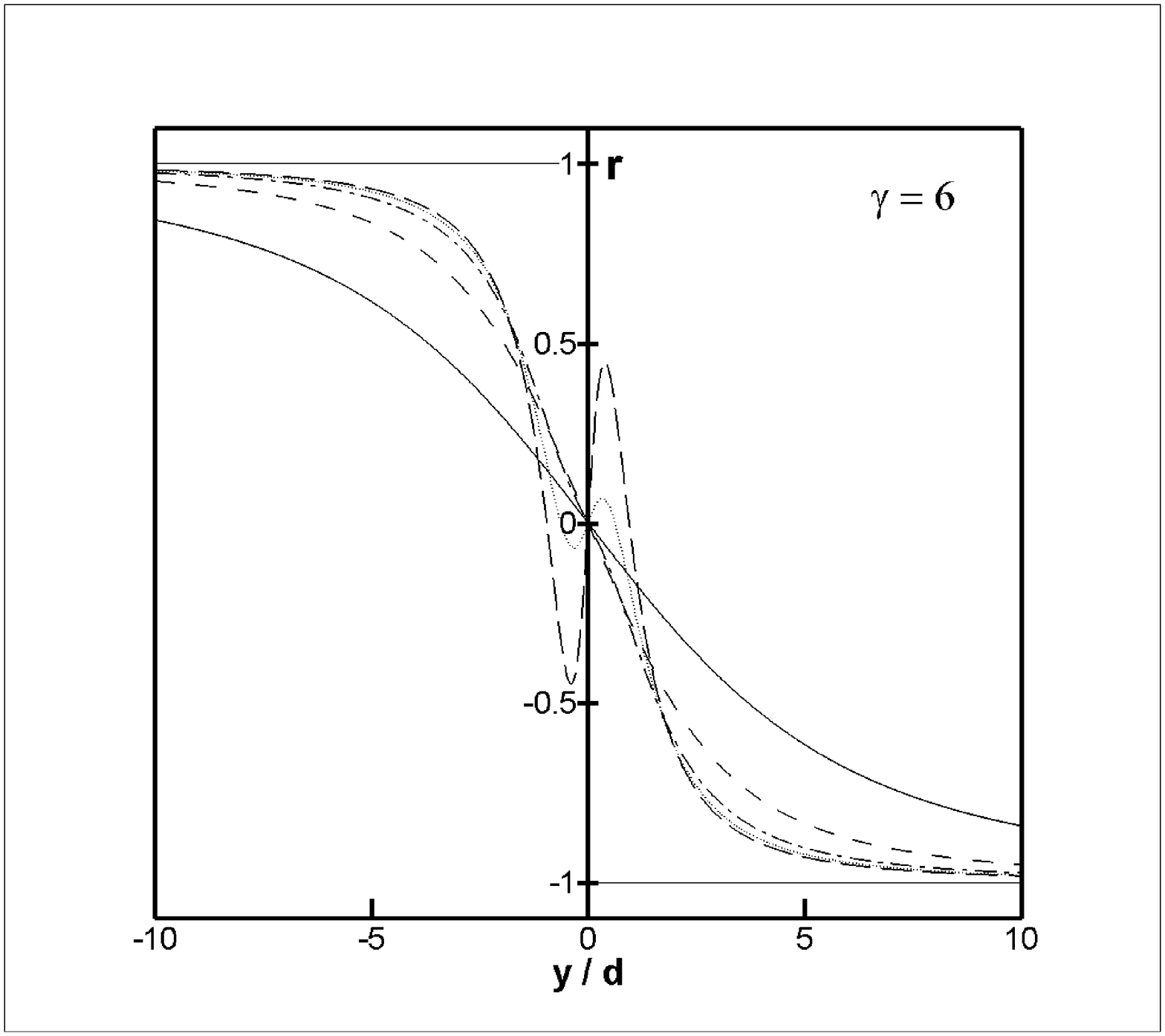}}
\caption{Profiles of strain persistence parameter for density
         ratio $ \gamma = 6 $.
         Solid: $ x/d = 0.25 $;
         dashed: $ x/d = 0.50 $;
         dashdot: $ x/d = 0.75 $;
         dotted: $ x/d = 0.90 $;
         longdash: $ x/d = 1.0 $.}
\label{Fig:fig4}
\end{center}
\end{figure}

\newpage

\begin{figure}[htpb]
\begin{center}
\scalebox{0.5}{\includegraphics[angle=-90]{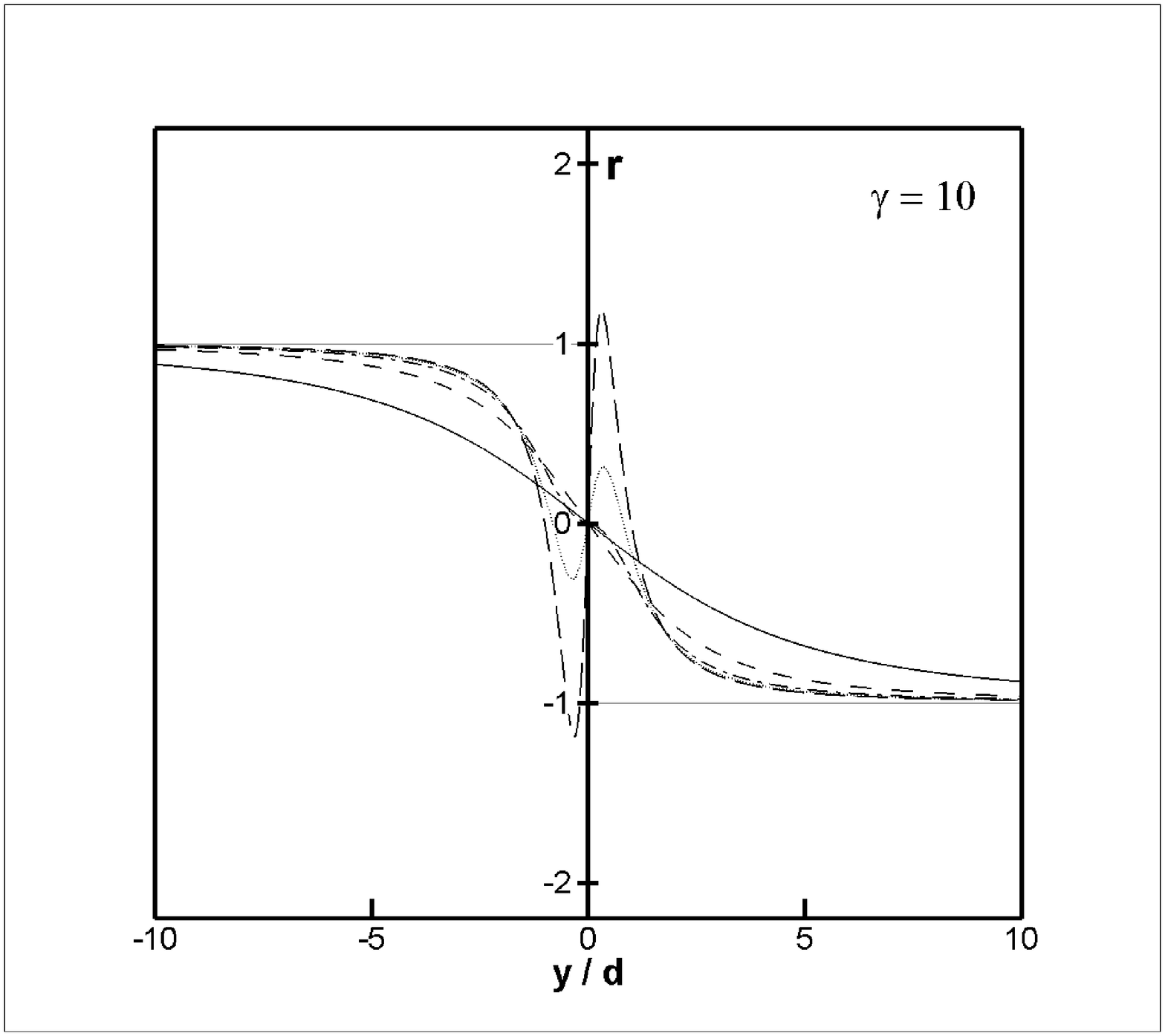}}
\caption{Profiles of strain persistence parameter for density
         ratio $ \gamma = 10 $.
         Solid: $ x/d = 0.25 $;
         dashed: $ x/d = 0.50 $;
         dashdot: $ x/d = 0.75 $;
         dotted: $ x/d = 0.90 $;
         longdash: $ x/d = 1.0 $.}
\label{Fig:fig5}
\end{center}
\end{figure}

\newpage

\begin{figure}[htpb]
\begin{center}
\scalebox{0.5}{\includegraphics[angle=-90]{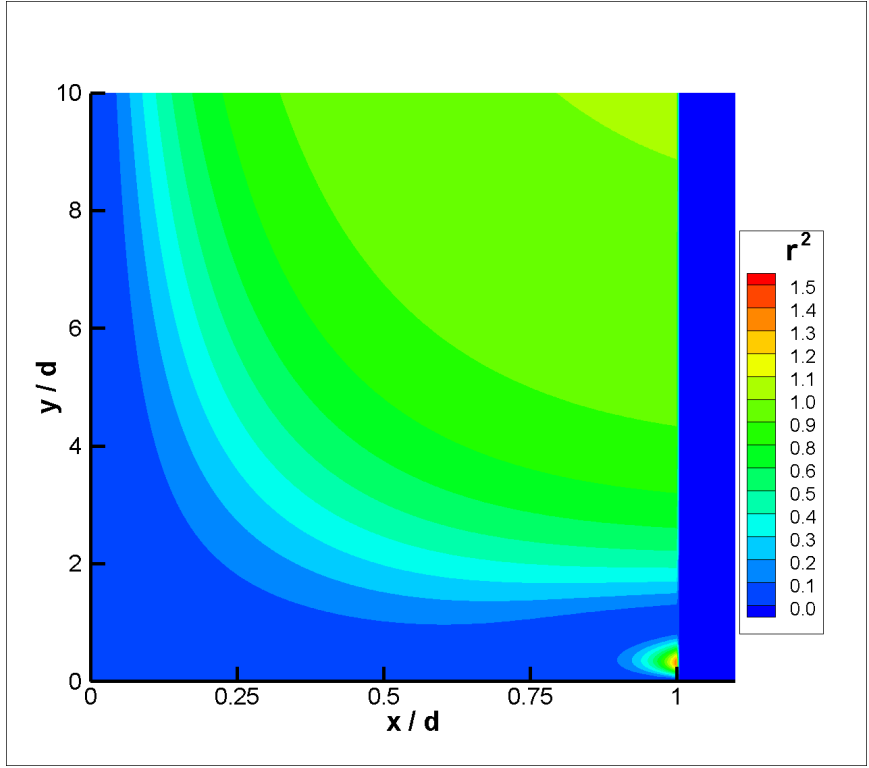}}
\caption{Field of strain persistence parameter -- squared --
         for density ratio $ \gamma = 10 $.}
\label{Fig:fig6}
\end{center}
\end{figure}

\newpage

\begin{figure}[htpb]
\begin{center}
\scalebox{0.5}{\includegraphics[angle=-90]{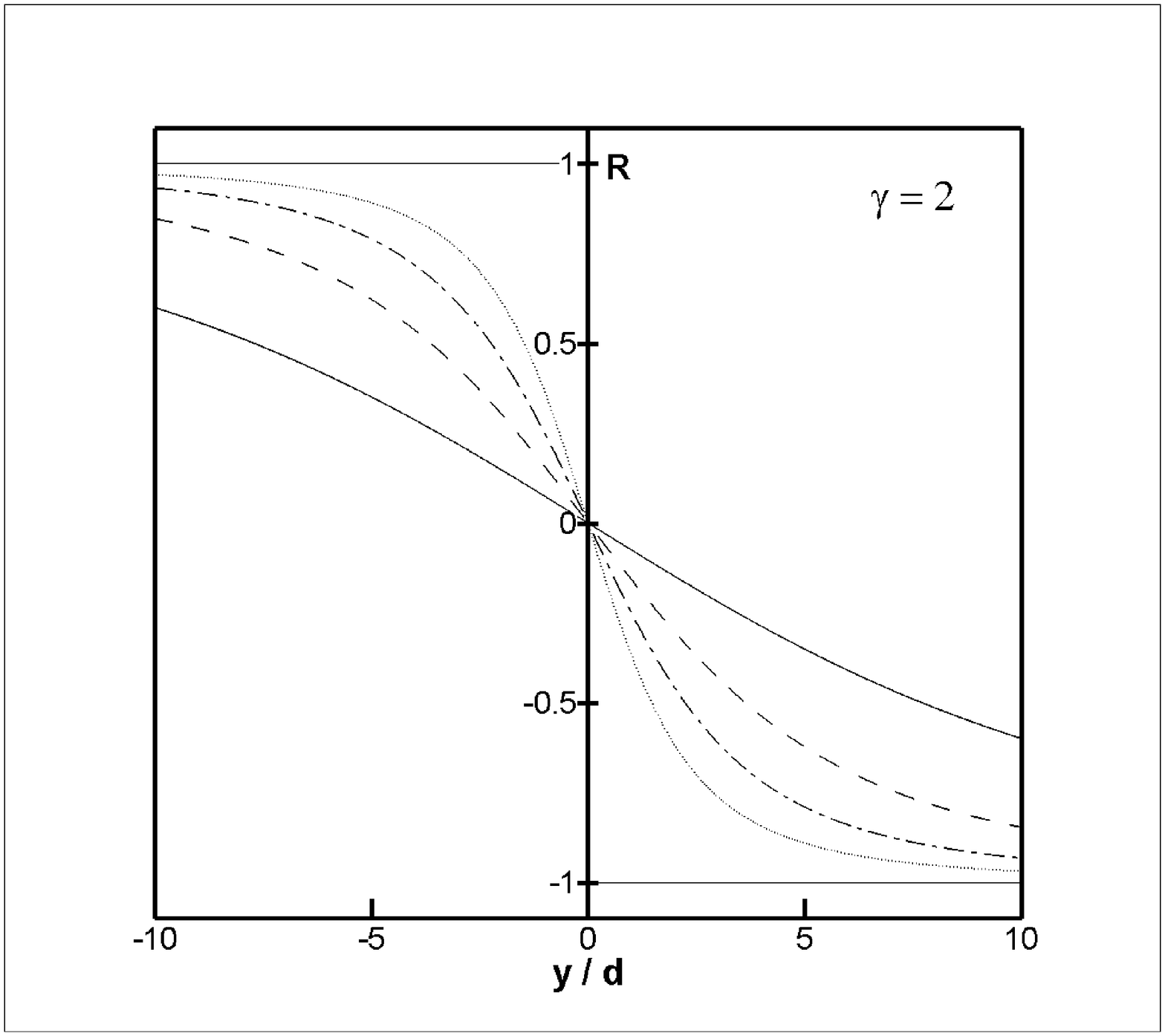}}
\caption{Profiles of vorticity-to-strain ratio for density
         ratio $ \gamma = 2 $.
         Solid: $ x/d = 0.25 $;
         dashed: $ x/d = 0.50 $;
         dashdot: $ x/d = 0.75 $;
         dotted: $ x/d = 1.0 $.}
\label{Fig:fig7}
\end{center}
\end{figure}

\newpage

\begin{figure}[htpb]
\begin{center}
\scalebox{0.5}{\includegraphics[angle=-90]{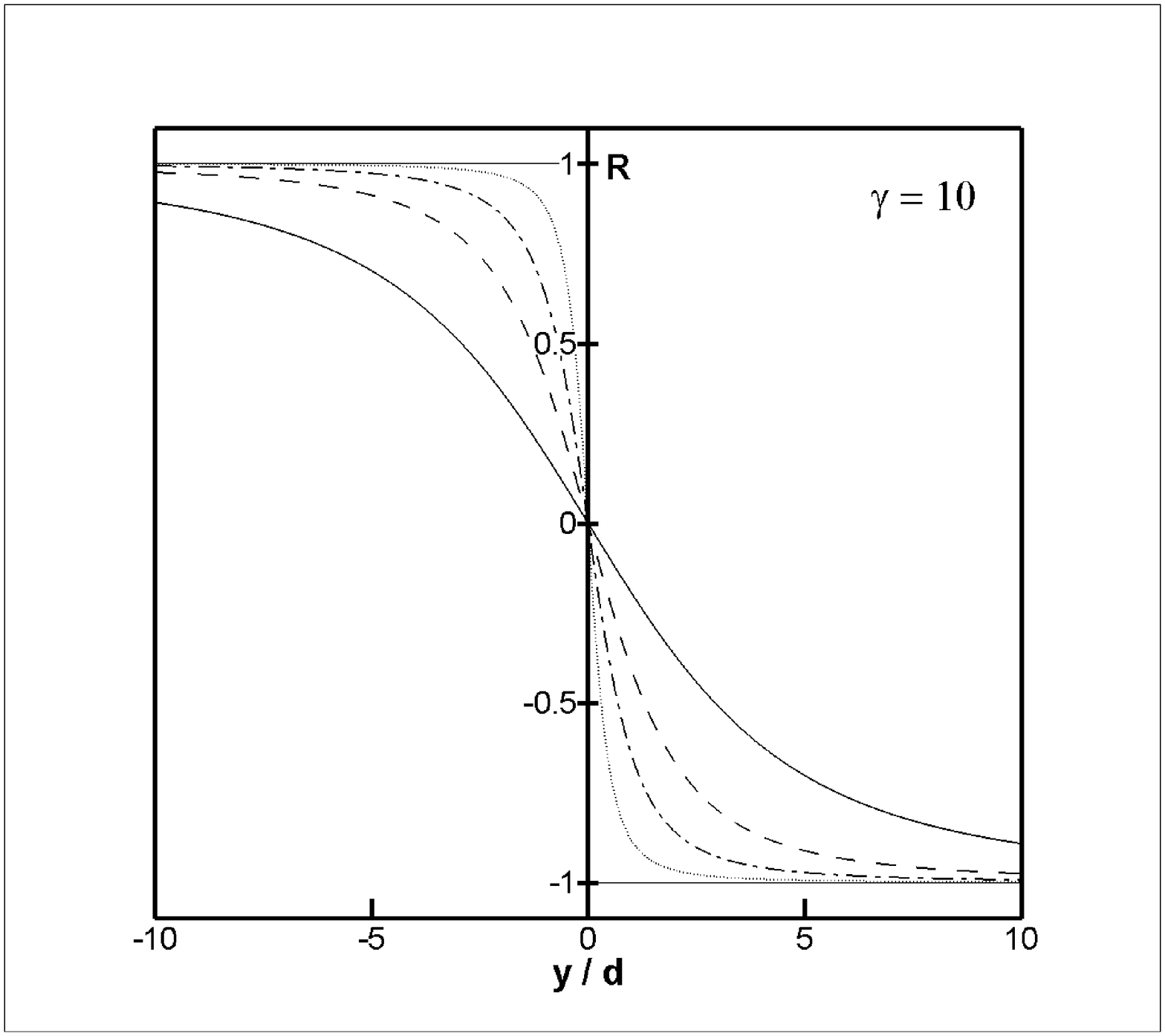}}
\caption{Profiles of vorticity-to-strain ratio for density
         ratio $ \gamma = 10 $.
         Solid: $ x/d = 0.25 $;
         dashed: $ x/d = 0.50 $;
         dashdot: $ x/d = 0.75 $;
         dotted: $ x/d = 1.0 $.}
\label{Fig:fig8}
\end{center}
\end{figure}

\newpage

\begin{figure}[htpb]
\begin{center}
\scalebox{0.5}{\includegraphics[angle=-90]{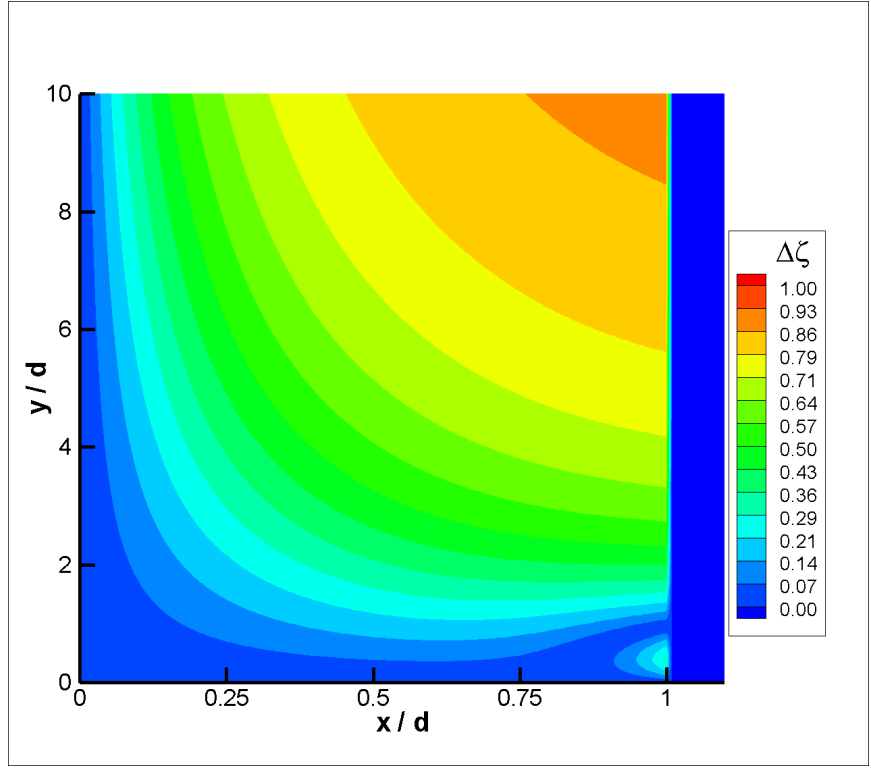}}
\caption{
Normalised difference between the local compressional
direction and the equilibrium orientation of the gradient
of a passive scalar for density ratio $ \gamma = 6 $.}
\label{Fig:fig9}
\end{center}
\end{figure}

\newpage

\begin{figure}[htpb]
\begin{center}
\scalebox{0.5}{\includegraphics[angle=-90]{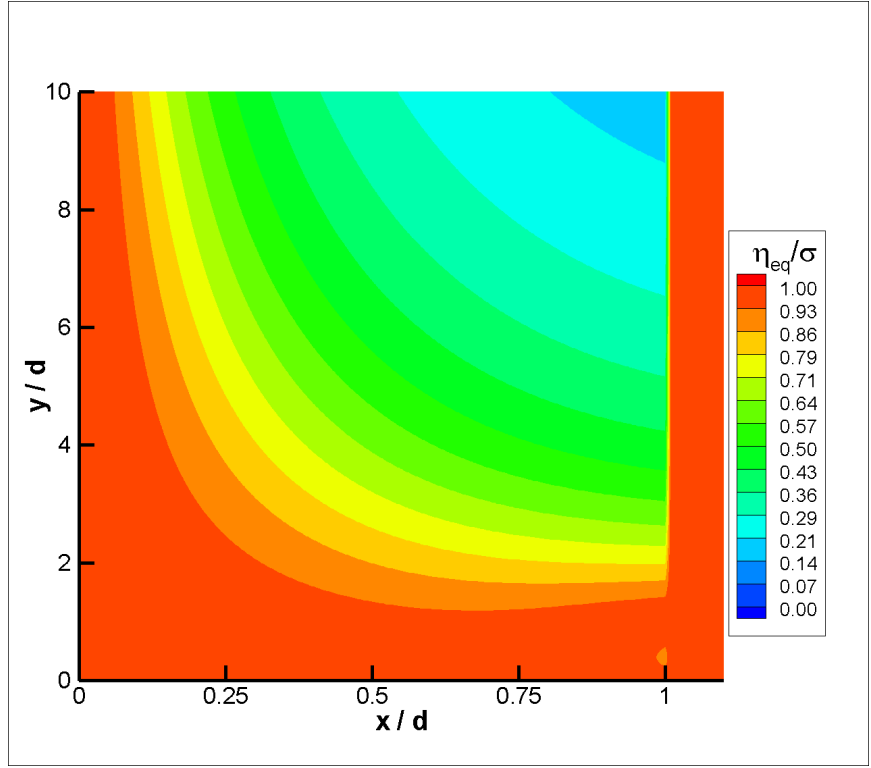}}
\caption{
Theoretical growth rate of the norm of a passive scalar
gradient 
normalised by local strain;
density ratio $ \gamma = 6 $.}
\label{Fig:fig10}
\end{center}
\end{figure}

\newpage

\begin{figure}[htpb]
\begin{center}
\scalebox{0.5}{\includegraphics[angle=-90]{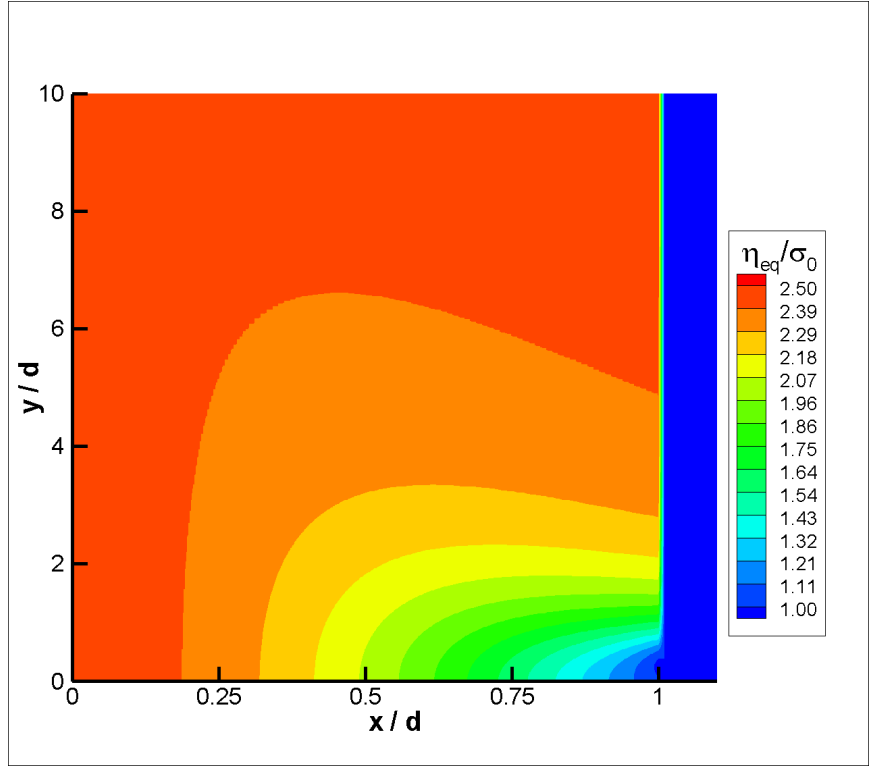}}
\caption{
Theoretical growth rate of the norm of a passive scalar
gradient 
normalised by the imposed strain;
density ratio $ \gamma = 6 $.}
\label{Fig:fig11}
\end{center}
\end{figure}

\newpage

\begin{figure}[htpb]
\begin{center}
\scalebox{0.5}{\includegraphics{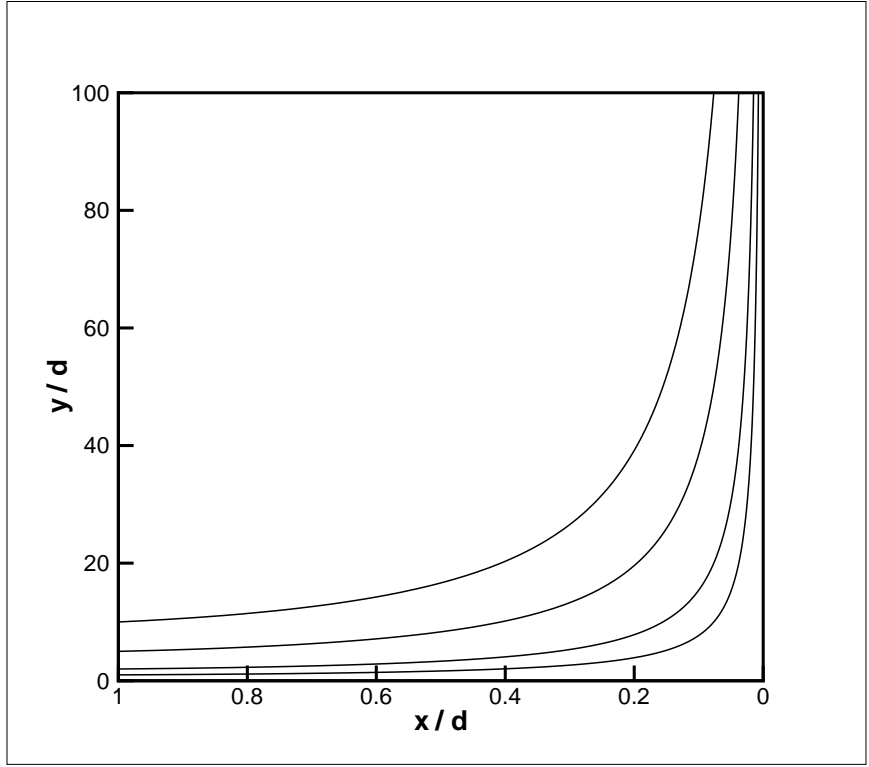}}
\caption{Trajectories starting from $ x/d = 1 $
and $ y/d = $ 1, 2, 5  and 10; density ratio $ \gamma = 6 $.}
\label{Fig:fig12}
\end{center}
\end{figure}

\newpage

\begin{figure}[htpb]
\begin{center}
\scalebox{0.5}{\includegraphics{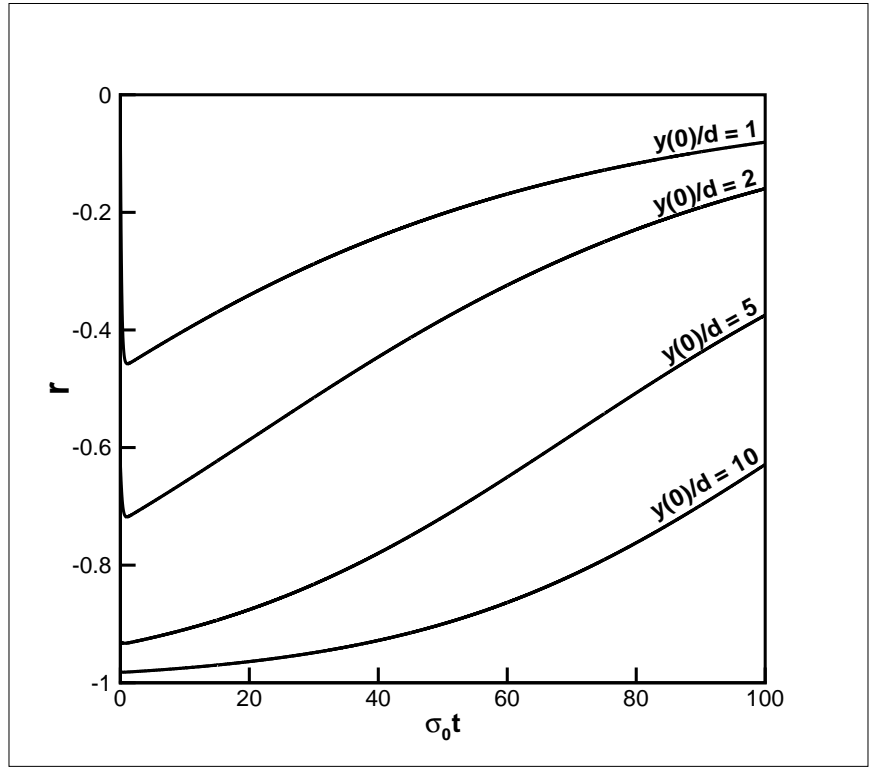}}
\caption{Evolution of strain persistence parameter
along trajectories starting from $ x/d =1 $
and $ y/d = $ 1, 2, 5  and 10; density ratio $ \gamma = 6 $.}
\label{Fig:fig13}
\end{center}
\end{figure}

\newpage

\begin{figure}[htpb]
\begin{center}
\scalebox{0.5}{\includegraphics{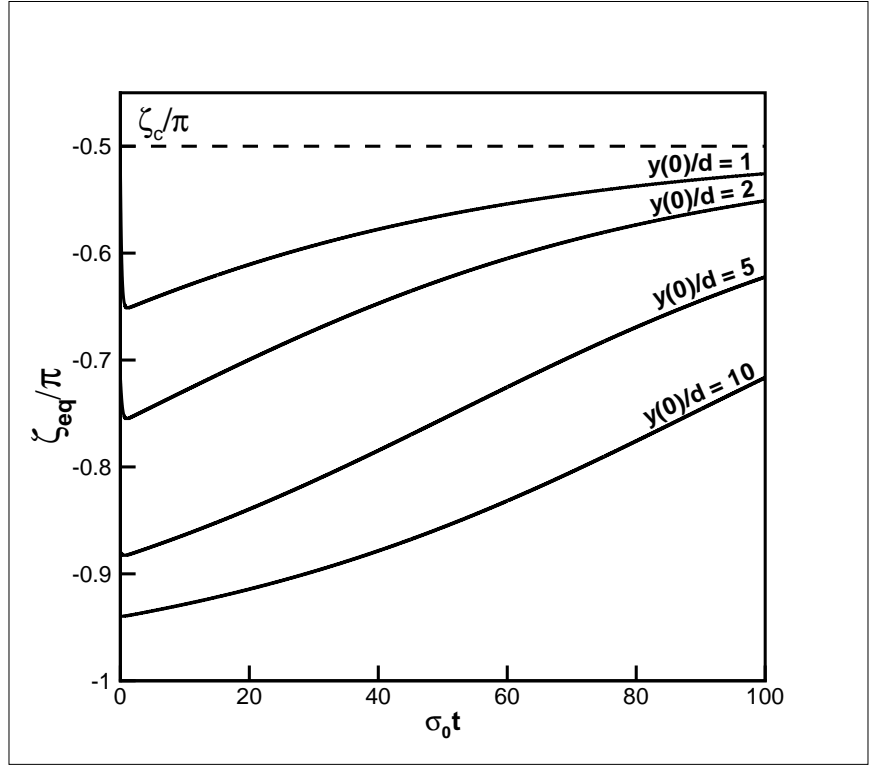}}
\caption{Evolution of equilibrium orientation
along trajectories starting from $ x/d =1 $
and $ y/d = $ 1, 2, 5  and 10; density ratio $ \gamma = 6 $.}
\label{Fig:fig14}
\end{center}
\end{figure}

\newpage

\begin{figure}[htpb]
\begin{center}
\scalebox{0.5}{\includegraphics{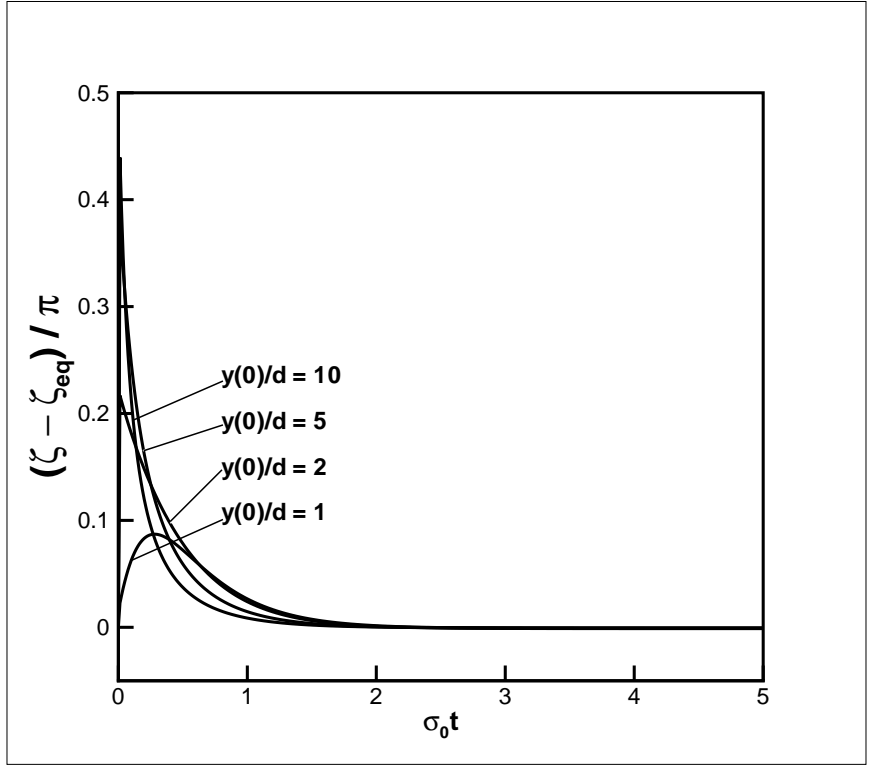}}
\caption{Departure of scalar gradient orientation to
equilibrium orientation
along trajectories starting from $ x/d =1 $
and $ y/d = $ 1, 2, 5  and 10; density ratio $ \gamma = 6 $.}
\label{Fig:fig15}
\end{center}
\end{figure}

\newpage

\begin{figure}[htpb]
\begin{center}
\scalebox{0.5}{\includegraphics{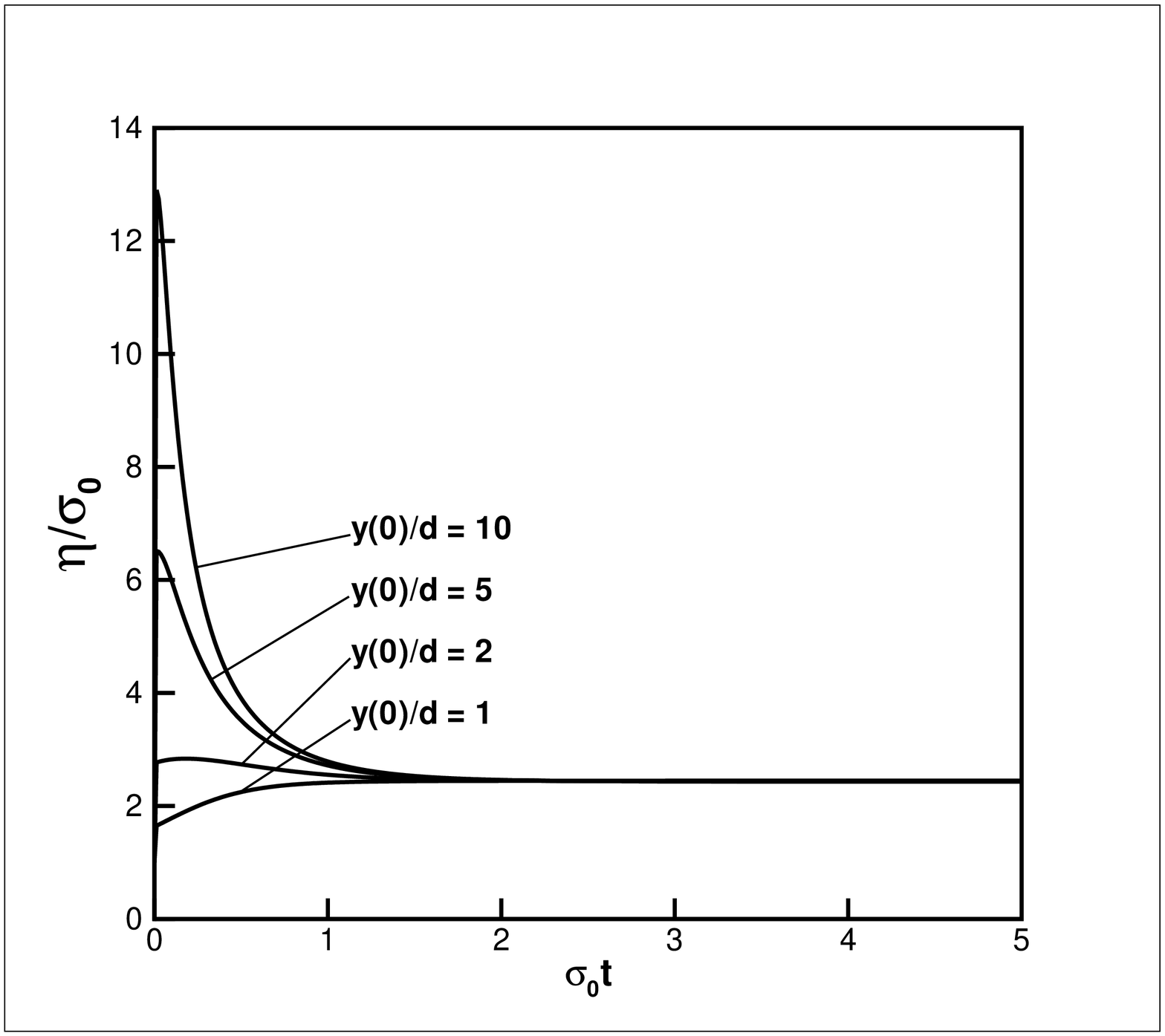}}
\caption{
Evolution of the actual growth rate of scalar gradient norm
normalised by the imposed strain
along trajectories starting from $ x/d =1 $
and $ y/d = $ 1, 2, 5  and 10; density ratio $ \gamma = 6 $.}
\label{Fig:fig16}
\end{center}
\end{figure}


\begin{thebibliography}{x}
\expandafter\ifx\csname natexlab\endcsname\relax\def\natexlab#1{#1}\fi
\expandafter\ifx\csname url\endcsname\relax
  \def\url#1{\texttt{#1}}\fi
\expandafter\ifx\csname urlprefix\endcsname\relax\def\urlprefix{URL }\fi

\bibitem[{Boratav {\it et al.} (1996)}]{Bal96}
Boratav, O.~N., Elghobashi, S.~E. and Zhong, R. 1996.
On the alignment of the $ \alpha $-strain and vorticity
in turbulent nonpremixed flames.
Phys. Fluids {\bf 8}, 2251-2253.

\bibitem[{Boratav {\it et al.} (1998)}]{Bal98}
Boratav, O.~N., Elghobashi, S.~E. and Zhong, R. 1998.
On the alignment of strain, vorticity and scalar gradient
in turbulent, buoyant, nonpremixed flames.
Phys. Fluids {\bf 10}, 2260-2267.

\bibitem[{Brethouwer {\it et al.} (2003)}]{Bal03}
Brethouwer, G., Hunt, J.~C.~R. and Nieuwstadt, F.~T.~M. 2003.
Micro-structure and Lagrangian statistics of the scalar field
with a mean gradient in isotropic turbulence.
J. Fluid Mech. {\bf 474}, 193--225.

\bibitem[{Chakraborty and Swaminathan (2007)}]{CS07}
Chakraborty, N. and Swaminathan, N. 2007.
Influence of the Damk\"ohler number on turbulence-scalar
interaction in premixed flames. I. Physical insight.
Phys. Fluids {\bf 19}, 045103.

\bibitem[{Chong {\it et al.} (1990)}]{Cal90}
Chong, M.~S., Perry, A.~E. and Cantwell, B.~J. 1990.
A general classification of three-dimensional flow fields.
Phys. Fluids {\bf 2}, 765--777.

\bibitem[{Diamessis and Nomura (2000)}]{DN00}
Diamessis, P.~J. and Nomura, K.~K. 2000.
Interaction of vorticity, rate-of-strain and scalar
gradient in stratified homogeneous sheared turbulence
Phys. Fluids {\bf 12}, 1166--1188.

\bibitem[{Dresselhaus and Tabor (1991)}]{DT91}
Dresselhaus, E. and Tabor, M. 1991.
The kinematics of stretching and alignments of material
elements in general flow fields.
J. Fluid Mech. {\bf 236}, 415--444.

\bibitem[{Garcia {\it et al.} (2005)}]{Gal05}
Garcia, A., Gonzalez, M. and Parantho\"en P. 2005.
On the alignment dynamics of a passive scalar gradient
in a two-dimensional flow.
Phys. Fluids {\bf 17}, 117102.

\bibitem[{Garcia {\it et al.} (2008)}]{Gal08}
Garcia, A., Gonzalez, M. and Parantho\"en P. 2008.
Nonstationary aspects of passive scalar gradient behaviour.
Eur. J. Mech. Fluids B/Fluids {\bf 27}, 433--443.

\bibitem[{Hua and Klein (1998)}]{HK98}
Hua, B.~L. and Klein, P. 1998. 
An exact criterion for the stirring properties of nearly
two-dimensional turbulence.
Physica D {\bf 113}, 98--110.

\bibitem[{Hua {\it et al.} (1998)}]{Hal98}
Hua, B.~L., McWilliams, J.~C.  and Klein, P. 1998. 
Lagrangian acceleration in geostrophic turbulence.
J. Fluid Mech. {\bf 366}, 87-108.

\bibitem[{Jackson and Matalon (1993)}]{JM93}
Jackson, T.~L. and Matalon, M. 1993. 
Stability of a premixed flame in stagnation point flow
against general disturbances.
Combust. Sci. Tech. {\bf 90}, 385--403.

\bibitem[{Kim and Matalon (1988)}]{KM88}
Kim, Y.~D. and Matalon, M. 1988. 
Propagation and extinction of a flame in a stagnation-point flow.
Combust. Flame {\bf 73}, 303--313.

\bibitem[{Kim and Matalon (1990)}]{KM90}
Kim, Y.~D. and Matalon, M. 1990. 
On the stability of near-equidiffusional strained premixed
flames.
Combust. Sci. Tech. {\bf 69}, 85--97.

\bibitem[{Lapeyre {\it et al.} (1999)}]{Lal99}
Lapeyre, G., Klein, P. and Hua B.~L. 1999.
Does the tracer gradient vector align with the strain
eigenvector in 2D turbulence?
Phys. Fluids {\bf 11}, 3729--3737.

\bibitem[{Nomura and Elghobashi (1993)}]{NE93}
Nomura, K.~K. and Elghobashi, S.~E. 1993.
The structure of inhomogeneous turbulence in variable
density nonpremixed flames.
Theoret. Comput. Fluid Dynamics {\bf 5}, 153--175.

\bibitem[{Nomura and Post (1998)}]{NP98}
Nomura, K.~K. and Post, G.~K. 1998.
The structure and dynamics of vorticity and rate of strain
in incompressible homogeneous turbulence.
J. Fluid Mech. {\bf 377}, 65--97.

\bibitem[{Pantano {\it et al.} (2003)}]{Pal03}
Pantano, C., Sarkar, S. and Williams, F.~A. 2003.
Mixing of a conserved scalar in a turbulent reaction shear
layer.
J. Fluid Mech. {\bf 481}, 291--328.

\bibitem[{Pumir} (1994)]{P94}
Pumir, A.
A numerical study of the mixing of a passive scalar
in three dimensions in the presence of a mean gradient.
Phys. Fluids {\bf 6}, 2118--2132.

\bibitem[{Vedula {\it et al.}} (2001)]{Val01}
Vedula, P., Yeung, P. K. and Fox, R.~O. 2001.
Dynamics of scalar dissipation in isotropic turbulence:
a numerical and modelling study.
J. Fluid Mech. {\bf 554}, 457--475.

\end{thebibliography}
\end{document}